\documentstyle[10lomcon,cite,amsmath,graphicx]{article}

\newcommand{\tr}{\mathop{\rm tr}\nolimits}

\newcommand{\Pexp}{\mathop{\rm Pexp}\nolimits}
\newcommand{\inst}{\mathop{\rm inst}\nolimits}
\newcommand{\T}{\mathop{\rm T}\nolimits}
\newcommand{\NP}{\mathop{\rm NP}\nolimits}
\newcommand{\PP}{\mathop{\rm P}\nolimits}
\newcommand{\eff}{\mathop{\rm eff}\nolimits}
\newcommand{\dia}{\mathop{\rm dia}\nolimits}
\newcommand{\para}{\mathop{\rm para}\nolimits}

\newcommand{\cool}{\mathop{\rm cool}\nolimits}

\begin{document}

\title{INSTANTON IN THE NONPERTURBATIVE QCD VACUUM}
\author{
N.O.~Agasian \footnote{e-mail: agasian@heron.itep.ru},
S.M.~Fedorov \footnote{e-mail: fedorov@heron.itep.ru}}

\address{
Institute of Theoretical and Experimental Physics,\\
117218, Moscow, B.Cheremushkinskaya 25, Russia}

\maketitle\abstracts{ The influence of nonperturbative
fields on instantons in quantum chromodynamics is studied.
Effective action for instanton is derived in bilocal approximation and it is
demonstrated that stochastic background gluon fields are
responsible for IR stabilization of instantons. It is shown that
instanton size in QCD is of order of 0.25~fm. Comparison of
obtained instanton size distribution with lattice data is made. }

\noindent{\bf 1.} Instantons  were introduced in 1975 by Polyakov
and coauthors~\cite{BPST}. These topologically nontrivial field
configurations are essential for the solution of some problems of
quantum chromodynamics. Instantons allow to explain anomalous
breaking of $U(1)_A$ symmetry and the $\eta'$
mass~\cite{tHooft_76b,Witten_Venez_79}, spontaneous chiral
symmetry breaking (SCSB)~\cite{Diak_Pet_86}. Taking into account
instantons is of crucial importance for many phenomena of QCD
(see~\cite{Scha_Shur_98} and references therein).

At the same time, there is a number of serious problems in
instanton physics. The first is IR inflation of instanton, i.e.
divergence of integrals over instanton size $\rho$ at big $\rho$.
Second, quasiclassical instanton anti-instanton vacuum lacks confinement.

The most popular model of instantons is the model of ''instanton liquid'',
which was phenomenologically formulated by Shuryak~\cite{Shuryak_81}. It states that average
distance between pseudoparticles is $\bar{R}\sim 1$~fm and their
average size is $\bar{\rho}\sim 1/3$~fm. Thus,
$\bar{\rho}/\bar{R}\simeq 1/3$ and vacuum consists of well
separated, and therefore not very much deformed, instantons and
anti-instantons. However, the mechanism for the
suppression of large-size instantons in the ensemble of
topologically non-trivial fields is still not understood.

On the other hand, QCD vacuum contains not only quasiclassical
instantons, but other nonperturbative fields as well. In this talk
we will demonstrate that instanton can be stabilized in
nonperturbative vacuum and exist as a stable topologically
nontrivial field configuration against the background of
stochastic nonperturbative fields, which are responsible for
confinement, and will find quantitatively it's size. In this way,
we will follow the analysis performed in~\cite{AF_hep}.

\noindent {\bf 2.}
Standard euclidian action of gluodynamics has
the form
\begin{equation}
\label{eq_action} S[A]=\frac{1}{2g_0^2} \int d^4 x
\tr(F_{\mu\nu}^2[A])= \frac{1}{4} \int d^4 x
F_{\mu\nu}^a[A]F_{\mu\nu}^a[A],
\end{equation}
where $ F_{\mu\nu}[A]=\partial_{\mu}A_{\nu} -
\partial_{\nu}A_{\mu}-i[A_{\mu},A_{\nu}]$ is the strength of gluon
field and we use the Hermitian matrix form for gauge fields
$A_\mu(x)= g_0A^a_\mu(x) {t^a}/{2}$ and $\tr
t^at^b=\delta^{ab}/2$.
We decompose $A_{\mu}$ as $A_{\mu} = A_{\mu}^{\inst}+B_{\mu}+a_{\mu}$,
where $A_{\mu}^{\inst}$ is an instanton-like field configuration
with a unit topological charge $Q_{\T}[A^{\inst}]=1$; $a_{\mu}$ is
quantum field and $B_{\mu}$ is nonperturbative background field (with zero
topological charge), which can be parametrized by gauge invariant
nonlocal vacuum averages of gluon field strength\footnote{In
operator product expansion method and in QCD sum rules
nonperturbative field is characterized by a set of local gluon
condensates $\langle G^2\rangle$, $\langle G^3\rangle$, \ldots}.

In general case effective action for instanton in NP vacuum takes
the form
\begin{equation}
\label{eq_genrl} Z=e^{-S_{\eff}[A^{\inst}]} = \int
[Da_{\mu}]\left\langle e^{-S[A^{\inst}+B+a]}\right\rangle,
\end{equation}
where $\langle...\rangle$ implies averaging over background field
$B_{\mu}$.

Integration over $a_{\mu}$ and $B_{\mu}$ corresponds to averaging over fields
that are responsible for the physics at different scales.
Integration over $a_{\mu}$ takes into account perturbative gluons
and describes phenomena at small distances, while averaging over
$B_{\mu}$ (formally interaction with gluon condensate) accounts
for phenomena at scales of confinement radius. Therefore averaging
factorizes $Z = \langle Z_1 \rangle \langle Z_2 \rangle$
(see~\cite{AF_hep} for details) and effective action appears to be
sum of two terms, ''perturbative'' and ''nonperturbative''.
Perturbative fluctuations were considered in~\cite{Agas_Sim_95,
Agasian_96}, and it was shown that in NP vacuum standard
perturbation theory for instantons changes, which results in
''freezing'' of effective coupling constant. The perturbative part
of effective instanton action in stochastic vacuum
$S^{\PP}_{\eff}[A^{\inst}]$ was shown to be
\begin{equation}
\label{eq_seff_pt_final} S^{\PP}_{\eff}(\rho)=\frac{b}{2}
\ln\frac{1/\rho^2+m_*^2}{\Lambda^2}
\end{equation}
Here $m_*\simeq 0.75 m_{0^{++}}  \sim 1\mbox{GeV}$, where $0^{++}$
is the lightest glueball.

Thus, we have for effective instanton action
\begin{align}
\label{eq_seff_gnrl}
&S_{\eff}[A^{\inst}]=S^{\PP}_{\eff}[A^{\inst}]+S^{\NP}_{\eff}[A^{\inst}]\\
&S^{\NP}_{\eff}[A^{\inst}]=-\ln\langle
Z_2(B)\rangle= -\ln\left\langle \exp
\{-S[A^{\inst}+B]+S[A^{\inst}]\}\right\rangle
\end{align}

\noindent {\bf 3.}
We consider effect of NP fields on instanton, i.e. we evaluate
$\langle Z_2\rangle$. In this work we make use of the method of
vacuum correlators, introduced in works of Dosch and
Simonov~\cite{Dosch_87}. NP vacuum of QCD is described in terms of
gauge invariant vacuum averages of gluon fields (correlators)
$$
\Delta_{\mu_1 \nu_1 ... \mu_n \nu_n}=\langle\tr
G_{\mu_1\nu_1}(x_1) \Phi(x_1,x_2) G_{\mu_2\nu_2}(x_2) ...
G_{\mu_n\nu_n}(x_n) \Phi(x_n,x_1)\rangle,
$$
where $G_{\mu\nu}$ is gluon field strength, and
$\Phi(x,y)=\Pexp\left(i\int\limits_y^x B_{\mu}dz_{\mu}\right)$ is
a parallel transporter, which ensures gauge invariance. In many
cases bilocal approximation appears to be sufficient for
qualitative and quantitative description of various physical
phenomena in QCD. Moreover, there are indications that corrections
due to higher correlators are small~\cite{DShS}.

Tensor structure of bilocal correlator follows from
antisymmetry in Lorentz indices. It is parametrized by two functions $D(x-y)$ and $\overline{D}(x-y)$:
\begin{align}
\label{eq_bilocal} &\langle g^2 G_{\mu\nu}^a(x,x_0)
G_{\rho\sigma}^b(y,x_0)\rangle
= \langle G^2 \rangle \frac{\delta^{ab}}{N_c^2-1} \times \notag\\
&\quad \times\left\{
\frac{D(z)}{12}\delta_{\mu\rho}\delta_{\nu\sigma}
+\frac{\overline{D}(z)}{6}(n_{\mu}n_{\rho}\delta_{\nu\sigma}+n_{\nu}n_{\sigma}\delta_{\mu\rho}) -
(\mu \leftrightarrow \nu)
\right\},
\end{align}
where $G_{\mu\nu}(x,x_0)=\Phi(x_0,x)G_{\mu\nu}(x)\Phi(x,x_0)$,~~
$n_{\mu}={z_{\mu}}/{|z|}={(x-y)_{\mu}}/{|x-y|}$ is the unit
vector, $\langle G^2 \rangle \equiv \langle g^2 G_{\mu\nu}^a
G_{\mu\nu}^a \rangle$ and, as it follows from normalization,
$D(0)+\overline{D}(0)=1$.

Bilocal correlator was measured on the lattice
(see~\cite{DiGiacomo_2000} and references therein), and functions
$D(z)$ and $\overline{D}(z)$ were found to be exponentially
decreasing $D(z)=A_0 \exp(-z/T_g)$, $\overline{D}(z)=A_1 z
\exp(-z/T_g)/T_g$, where $T_g$ is the gluonic correlation length.
Besides, according to lattice measurements $A_1 \ll A_0$ ($A_1\sim
A_0/10$). Lattice data are presented in Table~\ref{tab_digiacomo}.
$SU(3)$~full stands for chromodynamics with 4 quarks, while
$SU(2)$ and $SU(3)$~quenched mean pure $SU(2)$ and $SU(3)$
gluodynamics, respectively.

\begin{table}[!htb]
\caption{Lattice data~\cite{DiGiacomo_2000} for bilocal
correlator} \label{tab_digiacomo}
\parbox{2.5cm}{~}
\begin{tabular}{|lcc|}
\hline
& $\langle G^2\rangle$, GeV$^4$ & $T_g$,~fm\\
\hline
$SU(2)$ quenched & 13 & 0.16 \\
$SU(3)$ quenched & 5.92 & 0.22 \\
$SU(3)$ full & 0.87 & 0.34\\
\hline
\end{tabular}
\end{table}

To evaluate $S_{\eff}^{\NP}$ we
use the cluster expansion:
\begin{equation}
\langle \exp(x) \rangle =
 \exp \left( \langle x\rangle + \frac{\langle x^2 \rangle - \langle x \rangle ^2}{2!} + \ldots \right)
\end{equation}
In bilocal approximation we find
$S_{\eff}^{\NP}=S_{\dia}+\frac{1}{2}S_{\dia}^2+S_{\para}+S_1+S_2$,
where
\begin{align}
&S_{\dia}=-\frac{1}{2g^2}\int d^4 x
\left\langle\tr\left[\left([A_{\mu},B_{\nu}]-[A_{\nu},B_{\mu}]\right)^2\right]\right\rangle \label{eq_sdia}\\
&S_{\para}=-\frac{1}{2g^4}\int d^4 x d^4 y \left\langle
\tr\left(F_{\mu\nu}(x)G_{\mu\nu}(x)\right)
\tr\left(F_{\rho\sigma}(y)G_{\rho\sigma}(y)\right) \right\rangle \label{eq_spara}\\
&S_{1}=\frac{2}{g^4}\int d^4 x d^4 y \left\langle \tr\left(
F_{\mu\nu}[A_{\mu},B_{\nu}]\right)_x \tr\left(
F_{\rho\sigma}[A_{\rho},B_{\sigma}]
\right)_y \right\rangle  \label{eq_s1}\\
&S_{2}=\frac{2i}{g^4}\int d^4 x d^4 y\left\langle
\tr\left(F_{\mu\nu}G_{\mu\nu}\right)_x
\tr\left(F_{\rho\sigma}[A_{\rho},B_{\sigma}]\right)_y
\right\rangle \label{eq_s2}
\end{align}
We use notations $S_{\dia}$ (diamagnetic) and $S_{\para}$
(paramagnetic) for contributions~(\ref{eq_sdia})
and~(\ref{eq_spara}) into interaction of instanton with background
field. Next, $S_{\eff}^{\NP}$ can be expressed through bilocal correlator
(see~\cite{AF_hep} for details), for instance
\begin{align}
S_{\dia} = \frac{\langle G^2 \rangle}{12} \frac{N_c}{N_c^2-1} \int d^4 x
&\int\limits_0^1 \alpha d \alpha \int\limits_0^1 \beta d
\beta \, x^2 (A_{\mu}^a(x))^2 \times\\
&\times [D((\alpha-\beta)x)+2\overline{D}((\alpha-\beta)x)]
\end{align}

\noindent {\bf 4.}
We use standard form for instanton field configuration
\begin{equation}
\label{eq_inst}
A^{\inst}_{\mu} = 2 t^{b} R^{b \beta}
\overline{\eta}^{\beta}_{\mu\nu}
\frac{(x-x_0)_{\nu}}{(x-x_0)^2} f\left(\frac{(x-x_0)^{2}}{\rho^2}\right),
\end{equation}
where matrix $R^{b \beta}$ ensures embedding of instanton into $SU(N_c)$ group,
$b = 1,2,\,.\,.\,.N_c^2-1$; $\beta = 1,2,3$,
$\overline{\eta}^{\alpha}_{\mu\nu}$ are 't~Hooft symbols. In singular gauge profile
function $f(z)$ satisfies boundary conditions $f(0)=1$, $f(\infty)=0$ and
the classical solution has the form $f(z)=1/(1+z^2)$. Of course, real instanton profile
in NP vacuum is different. The problem of asymptotic behavior of
instanton solution far from the center $|x|\gg\rho_c$ was
studied in detail in Refs.~\cite{Diak_Pet_84, MAK, Dor}. Our numerical analysis shows that
the value of $\rho_c$ is almost not affected by the asymptotic of classical instanton solution
provided that condensate $\langle G^2\rangle$ and correlation length $T_g$ have reasonable values.

\begin{figure}[!htb]
\parbox{3cm}{~}
\begin{picture}(202,115)
\put(0,10){\includegraphics[scale=0.61]{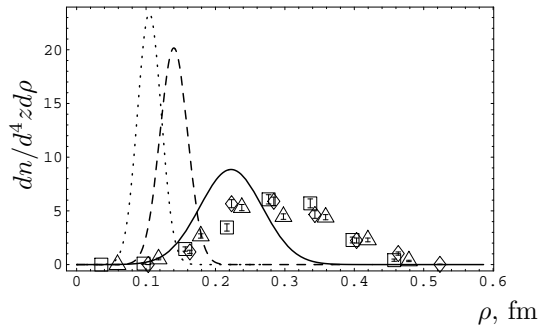}} \put(165,0){$\rho$, fm}
\put(-12,49){\rotatebox{90}{$dn/d^4 z d\rho$}}
\end{picture}
\caption{Instanton density $dn/d^4 z d\rho$ and lattice
data~\cite{Hasen_Niet_98}. $SU(3)$ full (solid line), $SU(2)$
quenched (dotted line)  and $SU(3)$ quenched (dashed line)}
\label{fig_2}
\end{figure}

Numerical calculations show that $S_{\dia}(\rho)$
is a growing function of $\rho$, and it ensured IR stabilization
of an instanton. Numerical results for instanton size distribution $dn/d^4 z d\rho \sim
\exp(-S_{\eff})$ and corresponding lattice
data~\cite{Hasen_Niet_98} are presented in Fig.~\ref{fig_2}. All
graphs are normalized to the commonly accepted instanton density
1~fm$^{-4}$. Different lattice groups roughly agree on
instantons size within a factor of two, e.g. $\bar{\rho}=0.3
\ldots 0.6$~fm for $SU(3)$ gluodynamics. There is no agreement at
all concerning the density $N/V$. As a tendency, lattice studies
give higher density and larger instantons than phenomenologically
assumed.

\begin{figure}[!htb]
\parbox{2.8cm}{~}
\begin{picture}(202,115)
\put(0,10){\includegraphics[scale=0.61]{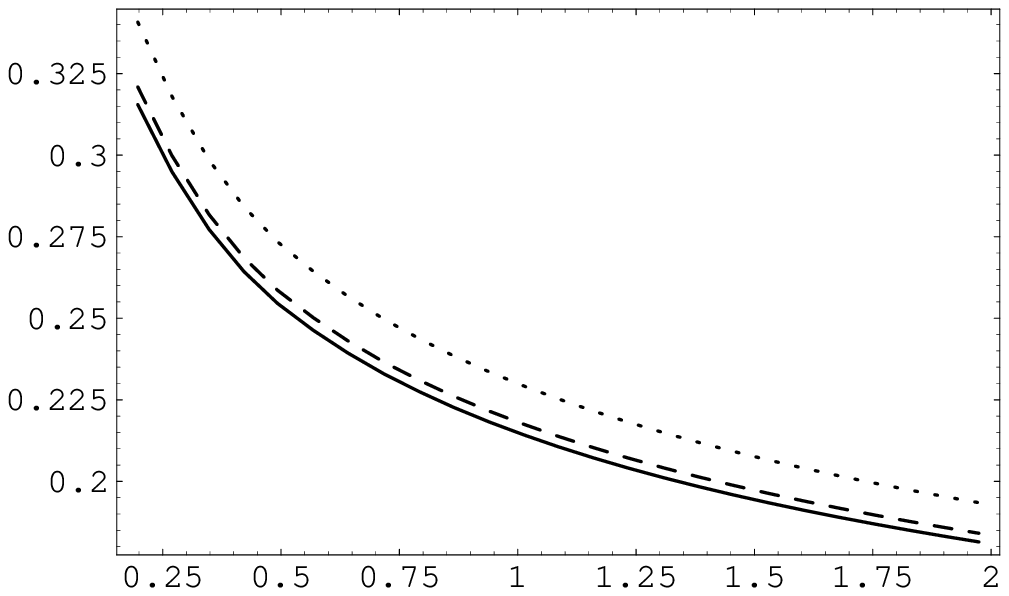}}
\put(165,0){$\langle G^2 \rangle$, GeV$^4$}
\put(-12,49){\rotatebox{90}{$\rho_c$, fm}}
\end{picture}
\caption{Instanton size as a function of gluon condensate ($N_c=3$, $N_f=4$) at
$T_g=0.2$~fm (dotted line), $T_g=0.3$~fm (dashed line), $T_g=0.34$~fm (solid line)}
\label{fig_3}
\end{figure}

Using our model we find for $\langle G^2\rangle = 5.92$~GeV$^4$
and $T_g=0.22$~fm that $\rho_c \simeq 0.15$~fm, which is less than
phenomenological ($\sim 0.3$~fm) and lattice results (in full
QCD we find $\rho_c \simeq 0.25$~fm).
However, we can present physical arguments to explain these deviations.
Lattice calculations include cooling procedure, during which some
lattice configurations of gluon field are discarded. This
procedure can result in a change in gluon condensate $\langle
G^2\rangle$, and thus instanton size distribution is calculated at
a value of gluon condensate $\langle G^2\rangle_{\cool}$ which is
smaller than physical value $\langle G^2\rangle$.
Therefore, lattice data for average instanton
size $\bar{\rho}$ should be compared with our calculations for
$\rho_c$, performed at smaller values of $\langle G^2\rangle$. We
show dependence of $\rho_c$ on $\langle G^2\rangle$ for several
values of $T_g$ in Fig.~\ref{fig_3}. One can see that increase of
$\langle G^2\rangle$ results in decrease of instanton size, and
that effect is a result of nonlocal ''diamagnetic'' interaction of
instanton with NP fields.

We did not go beyond bilocal approximation in this work. As
mentioned above, this approximation is good enough not only for
qualitative, but also for quantitative description of some phenomena in
nonperturbative QCD. In the problem under consideration there are
two small parameters. These are $1/g^2(\rho_c) \sim 0.15 \ldots
0.25$ and $1/N_c$, and there powers grow in each term of cluster
expansion. Moreover, we made an estimate for the sum of leading
terms in cluster expansion~\cite{AF}, and found that IR
stabilization stays intact ($\rho_c$ appears to be a little
smaller). Thus, proposed model describes physics of single
instanton stabilization in NP vacuum, not only qualitatively, but
also quantitatively with rather good accuracy.

\clearpage\newpage

\section*{Acknowledgments}
We are grateful to Yu.A.~Simonov for helpful discussions and comments. The
financial support of RFFI grant 00-02-17836 and INTAS grant CALL
2000 N 110 is gratefully acknowledged.


\section*{References}
\begin{thebibliography}{99}

\bibitem{BPST}
A.~M.~Polyakov,
%``Interaction Of Goldstone Particles In Two-Dimensions. Applications To Ferromagnets And Massive Yang-Mills Fields,''
{\em Phys.\ Lett.}\ {\bf B59} (1975) 79;~
%%CITATION = PHLTA,B59,79;%%
%
A.~A.~Belavin, A.~M.~Polyakov, A.~S.~Shvarts and Y.~S.~Tyupkin,
%``Pseudoparticle Solutions Of The Yang-Mills Equations,''
{\em Phys.\ Lett.}\ {\bf B59} (1975)~85.
%%CITATION = PHLTA,B59,85;%%

\bibitem{tHooft_76b}
G.~'t~Hooft,
%``Symmetry Breaking Through Bell-Jackiw Anomalies,''
{\em Phys.\ Rev.\ Lett.}\  {\bf 37} (1976) 8.
%%CITATION = PRLTA,37,8;%%

\bibitem{Witten_Venez_79}
E.~Witten,
%``Instantons, The Quark Model, And The 1/N Expansion,''
{\em Nucl.\ Phys.}\ {\bf B149} (1979) 285;~
%%CITATION = NUPHA,B149,285;%%
%
G.~Veneziano,
%``U(1) Without Instantons,''
{\em Nucl.\ Phys.}\ {\bf B159} (1979) 213.
%%CITATION = NUPHA,B159,213;%%

\bibitem{Diak_Pet_86}
D.~Diakonov and V.~Y.~Petrov,
%``A Theory Of Light Quarks In The Instanton Vacuum,''
{\em Nucl.\ Phys.}\ {\bf B272} (1986) 457.
%%CITATION = NUPHA,B272,457;%%

\bibitem{Scha_Shur_98}
T.~Schafer and E.~V.~Shuryak,
%``Instantons in QCD,''
{\em Rev.\ Mod.\ Phys.}\  {\bf 70} (1998) 323 [hep-ph/9610451].
%%CITATION = HEP-PH 9610451;%%

\bibitem{Shuryak_81}
E.~V.~Shuryak,
%``The Role Of Instantons In Quantum Chromodynamics. 1. Physical Vacuum,''
{\em Nucl.\ Phys.}\ {\bf B203} (1982) 93.
%%CITATION = NUPHA,B203,93;%%

\bibitem{AF_hep}
N.~O.~Agasian and S.~M.~Fedorov,
%``Instanton IR stabilization in the nonperturbative confining vacuum,''
hep-ph/0111208.
%%CITATION = HEP-PH 0111208;%%

\bibitem{Agas_Sim_95}
N.~O.~Agasian and Yu.~A.~Simonov,
%``Instantons in the stochastic QCD vacuum,''
{\em Mod.\ Phys.\ Lett.}\ {\bf A10} (1995) 1755.
%%CITATION = MPLAE,A10,1755;%%

\bibitem{Agasian_96}
N.~O.~Agasian,
%``Instantons In The Nonperturbative Confining Vacuum,''
{\em Phys.\ Atom.\ Nucl.}\  {\bf 59} (1996) 297.
%[Yad.\ Fiz.\  {\bf 59N2} (1996) 317].
%%CITATION = PANUE,59,297;%%

\bibitem{Dosch_87}
H.~G.~Dosch,
%``Gluon Condensate And Effective Linear Potential,''
{\em Phys.\ Lett.}\ {\bf B190} (1987) 177;~
%%CITATION = PHLTA,B190,177;%%
%
H.~G.~Dosch and Yu.~A.~Simonov,
%``The Area Law Of The Wilson Loop And Vacuum Field Correlators,''
{\em Phys.\ Lett.}\ {\bf B205} (1988) 339;~
%%CITATION = PHLTA,B205,339;%%
%
Yu.~A.~Simonov,
%``Vacuum Background Fields In QCD As A Source Of Confinement,''
{\em Nucl.\ Phys.}\ {\bf B307} (1988) 512.
%%CITATION = NUPHA,B307,512;%%

\bibitem{DShS}
A.~Di~Giacomo, H.~G.~Dosch, V.~I.~Shevchenko and Yu.~A.~Simonov,
%``Field correlators in QCD: Theory and applications,''
hep-ph/0007223.
%%CITATION = HEP-PH 0007223;%%

\bibitem{DiGiacomo_2000}
A.~Di~Giacomo,
%``Topics in non-perturbative QCD,''
hep-lat/0012013.
%%CITATION = HEP-LAT 0012013;%%

\bibitem{Diak_Pet_84}
D.~Diakonov and V.~Y.~Petrov,
%``Instanton Based Vacuum From Feynman Variational Principle,''
{\em Nucl.\ Phys.}\ {\bf B245} (1984) 259.
%%CITATION = NUPHA,B245,259;%%

\bibitem{MAK}
A.~B.~Migdal, N.~O.~Agasian and S.~B.~Khokhlachev,
%``Effect Of Quantum Fluctuations On The Shape Of An Instanton,''
{\em JETP Lett.}\  {\bf 41} (1985) 497;~
%[Pisma Zh.\ Eksp.\ Teor.\ Fiz.\  {\bf 41} (1985) 405].
%%CITATION = JTPLA,41,497;%%
%
N.~O.~Agasian and S.~B.~Khokhlachev,
%``Instantons In Gauge Theories With Broken Scale Invariance. 1: Gluodynamics,''
%``Instantons In Gauge Theories With Broken Scale Invariance: Gauge Theories With Higgs Fields,''
{\em Sov.\ J.\ Nucl.\ Phys.}\  {\bf 55} (1992) 628,~633;~
%[Yad.\ Fiz.\  {\bf 55} (1992) 1116,~1126].
%%CITATION = SJNCA,55,628;%%
%%CITATION = SJNCA,55,633;%%
%
N.~O.~Agasian,
%``Effects of screening of the instantons in nonperturbative {QCD},''
hep-ph/9803252;~
%%CITATION = HEP-PH 9803252;%%
%
%N.~O.~Agasian,
%``Instanton screening in the nonperturbative gluodynamics,''
hep-ph/9904227.
%%CITATION = HEP-PH 9904227;%%

\bibitem{Dor}
A.~E.~Dorokhov, S.~V.~Esaibegian, A.~E.~Maximov and S.~V.~Mikhailov,
%``Nonlocal gluon condensate within a constrained instanton model,''
{\em Eur.\ Phys.\ J.}\ {\bf C13} (2000) 331
[hep-ph/9903450].
%%CITATION = HEP-PH 9903450;%%


\bibitem{Hasen_Niet_98}
A.~Hasenfratz and C.~Nieter,
%``Instanton content of the SU(3) vacuum,''
{\em Phys.\ Lett.}\ {\bf B439} (1998) 366 [hep-lat/9806026].
%%CITATION = HEP-LAT 9806026;%%

\bibitem{AF}
N.O.~Agasian and S.M.~Fedorov, ITEP-PH-6/2001

\end{thebibliography}
\end{document}